\documentclass[12pt, conference]{IEEEtran}
\usepackage[T2A]{fontenc}
\usepackage[russian,english]{babel}
\usepackage{amsmath,amssymb}
\usepackage{multirow}
\usepackage{graphicx}
\usepackage{booktabs}
\usepackage{float}
\usepackage[ruled,vlined]{algorithm2e}
\usepackage{afterpage}
\usepackage{url}
\usepackage{cite}
\usepackage{hyperref} 
\usepackage[sort,compress]{natbib} 
\usepackage{xcolor}

\title{Terastate-per-second QUBO Brute-Force on a Single GPU: A Matrix Prefix-Suffix Decomposition}

\author{
  \IEEEauthorblockN{
    Aleksandr Maltsev\thanks{1,2}, 
    Mikhail Remnev\thanks{1}, 
    Alexey Kapranov\thanks{1} 
    and 
    Ekaterina Krivtsova\thanks{1}
  }
  \IEEEauthorblockA{
  }
  \IEEEauthorblockA{
    \thanks{2}Correspondence to: A. Maltsev, Email: \texttt{pulkin@gmail.com}.
  }
}

\begin{document}

\maketitle
\begin{abstract} 
\textbf{This paper presents a parallel QUBO exhaustive search algorithm for dense matrices, based on a prefix-suffix decomposition and Gray code ordering.
The algorithm achieves O(1) per-state complexity: for the QUBO objective function computation only one arithmetic operation per state is performed.
An adjustable energy components cache size enables placement in the fastest available memory tier. 
This reduces memory bandwidth requirements to a negligible level and transforms the problem from memory-bound to compute-bound.
Our CUDA-based implementation achieves a state-of-the-art evaluation rate of $7.5\times10^{12}$ states per second on a single GPU, setting a new performance benchmark for the full-space-search subclass of exact solvers.}
\end{abstract}

\begin{IEEEkeywords}
Quantum computing, Ising model, Ising spin–glass, QUBO, Combinatorial optimization, GPU, CUDA, Exhaustive search, Bruteforce, Quantum annealers, HPC
\end{IEEEkeywords}


\section{Introduction}

The QUBO (Quadratic Unconstrained Binary Optimization) problem involves minimizing the function \begin{equation*}E(x) = \sum_{i,j} Q_{ij} x_i x_j,\end{equation*} where $x \in \{0,1\}^n$, and is one of the central problems in combinatorial optimization~\cite{veloxq2025,graycode2023,kochenberger2014}. 
This problem is equivalent to finding the ground state of the Ising model~\cite{veloxq2025,dwaveqa2021}, described by the Hamiltonian \begin{equation*}H(s) = \sum_{i,j} J_{ij} s_i s_j + \sum_i h_i s_i.\end{equation*}

A wide range of combinatorial optimization problems can be reduced to the QUBO formulation, which underlies its practical importance.
These include graph-theoretic problems (Max-Cut, graph coloring, maximum independent set, community detection in networks), logistic problems such as the traveling salesman problem (TSP) with $N^2$ variables for solving an $N$-city problem, and financial applications (portfolio optimization)~\cite{Glover2022,nakano2023,lucas2014,kochenberger2014}.
Machine learning tasks (inference in graphical models, restricted Boltzmann machines), industrial applications (optimization of metal heat treatment schedules), as well as physical simulations (molecular modeling and ground state search in spin glasses) are also worth mentioning~\cite{app15168847,veloxq2025,dwaveqa2021,graycode2023,tropical2021,qhd2024}. Furthermore, a systematic transformation from general 0/1 linear programming models to QUBO formulations is available~\cite{Glover2022}.

The practical significance of these formulations is further strengthened by the fact that the Ising Hamiltonian is naturally expressible in the language of quantum computation, where binary spin variables are directly associated with qubit degrees of freedom and pairwise couplings can be encoded as quantum interactions. As a result, QUBO and Ising representations provide a broadly applicable route for translating real-world optimization tasks into forms amenable to quantum algorithms and quantum hardware, including adiabatic and gate-based approaches~\cite{lucas2014}.

The main reason for the difficulty of QUBO problems lies in their NP-hardness~\cite{veloxq2025,dwaveqa2021,graycode2023,kochenberger2014}. 
Even with just one additional bit in problem size, the state space doubles, reaching $2^n$ possible configurations~\cite{graycode2023}. 
Without special graph structure, finding an exact solution in polynomial time is impossible~\cite{veloxq2025}. 
This complexity is worsened by the presence of numerous local minima in the energy landscape~\cite{simb2019,simb2021}. 

Fully-connected problems have non-zero $Q_{ij}$ elements for all variable pairs, creating maximum computational complexity~\cite{momenta2019}. 
For example, a fully-connected problem on 100,000 spins contains about 5 billion edges~\cite{simb2019}. 
Sparse problems contain connections only between nearest neighbors or on specific topologies, such as the D-Wave Chimera graph, where each node connects to 5--6 others~\cite{veloxq2025,dwaveqa2021}. 

\subsection{Exact Solution Methods} 

Branch and Bound is one of the most powerful classical tools for exactly solving NP-hard problems~\cite{Barahona1989,dwaveqa2021,liers2004}. 
The algorithm recursively partitions the solution space into subproblems and uses relaxations to obtain lower bounds on the cost of the optimal solution~\cite{dwaveqa2021,liers2004}. 
For QUBO problems, odd-cycle inequalities, semidefinite programming (SDP), and dynamic programming methods for graphs with bounded treewidth are used~\cite{Rendl2010,dwaveqa2021,liers2004}. 

The SDP-based Branch-and-Cut methodology demonstrates successful results in solving Chimera graph problems with guaranteed optimal solutions~\cite{dwaveqa2021,liers2004}. 
Despite its power, Branch-and-Bound is practically applicable only to problems up to $\approx$100 variables in the fully-connected case~\cite{fasterexactsparce2023,Hrga2021}. 
For sparse matrices, problems with 10,000 nodes can be solved~\cite{fasterexactsparce2023}.

While QUBO formulations offer universality for solving combinatorial optimization problems on quantum and quantum-inspired hardware, direct problem-specific approaches often prove more efficient for large-scale instances. A notable example is the exact solution of the Traveling Salesman Problem with 85,900 cities \cite{Applegate2007, Applegate2009}, achieved through specialized branch-and-cut methods implemented in the Concorde solver, demonstrating that tailored algorithms can handle problem sizes far beyond the reach of current QUBO-based approaches while providing rigorous certification of optimality.

Tensor network methods enable exact ground state energy computation by compressing the tensor network using tropical algebra, successfully applied to problems up to 1024 spins on the Chimera graph~\cite{tropical2021}. 
Industrial solvers CPLEX and Gurobi are standard tools using integer programming~\cite{benchmark2025,qhd2024}. 
Specialized exact solvers like BEIT for Chimera topology can handle problems up to 1024 variables with optimality guarantees~\cite{veloxq2025}. 

\subsection{Heuristic Methods}

Simulated Annealing remains a classical method but requires sequential variable updates for its core search trajectory~\cite{zhang2022,benchmark2025}, which limits its degree of parallelism.
Newer physics-inspired algorithms show advantages: Simulated Bifurcation enables simultaneous updates of all variables~\cite{simb2019,simb2021}, while Momentum Annealing achieves 250x speedup compared to Simulated Annealing (SA) on GPU for fully-connected problems on 100,000 spins~\cite{momenta2019}.

GPU-based solvers demonstrate superior performance for QUBO problems through massive parallelization, as exemplified by the DABS (Diverse Adaptive Bulk Search) framework that achieves 100\% success rate on benchmark instances including K2000 and scales efficiently across 8 NVIDIA A100 GPUs, solving problems with up to 10,000 variables in seconds—100-200 times faster than quantum annealers while maintaining guaranteed optimality \cite{nakano2023}.

A class of specialized hardware solvers is also developing, using physical processes to find the energy minimum: memristor crossbars with quantum-inspired parallel annealing (QPA), coherent Ising machines (CIM), bifurcation algorithms on FPGA and GPU, oscillator systems, and magnetic devices~\cite{mohseni2022,memristor2023,daisuke2019}. Classical CMOS implementations of fully connected problems include STATICA, a 512-spin annealing processor fabricated in 65-nm technology that enables parallel spin updates through stochastic cellular automata dynamics and achieves state-of-the-art performance on combinatorial optimization \cite{statica2021}. Coherent Ising machines deserve special mention, as they enable solutions to fully connected problems with up to 100,000 variables \cite{spincoherentising2021}, though not optimally.

Machine learning approaches, particularly graph neural networks and deep reinforcement learning, have demonstrated remarkable potential for solving large-scale QUBO and Ising model problems. Physics-inspired graph neural networks achieve near-optimal solutions for instances with up to one million variables, demonstrating superior scalability compared to quantum annealers while maintaining approximation ratios above 90\% for Max-Cut problems \cite{schuetz2022}, while deep reinforcement learning methods have successfully found ground states of 3D spin glass systems with up to 8,000 spins—requiring 540 times fewer initial configurations than parallel tempering to guarantee exact solutions for smaller systems \cite{fan2023}.
Importantly, both the Max-Cut and 3D spin glass problems are equivalent to QUBO formulations, as Max-Cut can be directly expressed as a QUBO.

\subsection{Quantum Annealers}

D-Wave quantum annealers use qubit architectures organized into Chimera, Pegasus, and Zephyr graphs~\cite{veloxq2025,dwaveqa2021}. Quantum computers offer potential advantages for QUBO solving due to quantum phenomena like superposition and quantum tunneling~\cite{Nielsen_Chuang_2010,dwaveqa2021}. 

However, current quantum annealers have significant limitations: limited qubit count (D-Wave Advantage has $\sim$5000 qubits, while classical GPUs can handle problems with millions of variables)~\cite{veloxq2025,dwaveqa2021,benchmark2025}, limited connectivity topology requiring complex minor embedding~\cite{dwaveqa2021}, presence of noise and other error sources~\cite{dwaveqa2021}, and uncertain time complexity for practically relevant problem sizes~\cite{veloxq2025,dwaveqa2021}. 

Direct comparisons show that classical exact methods often outperform modern quantum annealers~\cite{veloxq2025,dwaveqa2021,benchmark2025}. In a study comparing the D-Wave 2000Q quantum annealer with optimal classical Branch-and-Cut methods, classical methods found optimal solutions faster for most problems~\cite{dwaveqa2021}. 

A comprehensive comparison of D-Wave's modern hybrid quantum solver with classical methods showed quantum advantages for large, dense QUBO matrices, achieving relative accuracy of 0.013 and 6561x solution time reduction compared to the best classical solver for 10,000-variable problems~\cite{benchmark2025}. 

A recent study by Upadhyay and Jones~\cite{upadhyay2025} systematically compared quantum annealing (D-Wave), digital annealing (Fujitsu DA v4~\cite{fujitsuda}), and classical MIP (Mixed Integer Programming)/CP (Constraint Programming) solvers on two industrial problems. For mRNA  codon optimization, Gurobi proved fastest with linear scaling and 100\% optimality, while D-Wave Leap NL HQA (Nonlinear Hybrid Quantum Annealer) achieved optimality with comparable performance, outperforming Fujitsu DA v4, which had the worst performance. For extra-large proteins, Gurobi maintained optimal solutions, while NL HQA showed suboptimal performance. A critically important result was the identification of rank-1 dominance: guanine-cytosine content quadratic terms form a rank-1 QUBO matrix, enabling classical solvers to efficiently linearize the problem. For reaction network pathway analysis, classical MIP/CP solvers (Gurobi, CP-SAT) vastly outperformed quantum-inspired approaches, achieving optimality with minimal computational effort.

\subsection{Brute-force Approaches}
Brute-force, by definition, explores all $2^N$ possible states, making its complexity inherently exponential. Despite this, brute-force remains important.
It guarantees optimality~\cite{graycode2023,veloxq2025} and serves as a benchmark for other methods~\cite{omnisolver2023,graycode2023,mehta2022}. Also, it is often faster than heuristics for 30 variable problems~\cite{worktime2020} and acceptable for 50 variables~\cite{bruteforcespin2021}, requires no parameter tuning~\cite{graycode2023}, and provides the best scalability on GPU clusters~\cite{mehta2022,bruteforcespin2021}, because traversing the state space is an "embarrassingly parallel" task. The entire state space can be partitioned into independent subsets processed by GPU cores simultaneously, without complex inter-node interaction \cite{worktime2020,jałowiecki2021}.
Additionally, brute-force enables not only finding the ground state but also constructing the low-energy spectrum~\cite{bruteforcespin2021}, which is highly useful for quantum algorithms~\cite{zhukov2025}.
Moreover, as noted in \cite{worktime2020}, the accuracy of solutions obtained by heuristic algorithms and quantum annealers depends on the type of problem, unlike that of brute-force.

Brute-force in its naive form has $O(N^2 \cdot 2^N)$ complexity, as computing energy for each of $2^N$ states takes $O(N^2)$ time~\cite{worktime2020}.

However, optimized brute-force algorithms significantly improve this by using Gray code to traverse the state space, enabling incremental energy updates in $O(N)$ time based on the previous value, reducing total complexity to $O(N \cdot 2^N)$~\cite{ graycode2023}.

Nevertheless, algorithms with $O(1)$ per state complexity are not novel. 
For such complexity, the evaluation rate in states/second is independent of problem size, making it a more practical metric than raw execution time.
In the paper~\cite{worktime2020}, constant-time energy updates are achieved by combining a tree-based depth-first traversal of the search space with Gray code ordering.
This enables evaluation speeds reaching up to $0.07 \times 10^{12}$ states per second on a single NVIDIA GeForce RTX 2080Ti GPU.

\begin{figure}[htbp]
    \centering
    \includegraphics[width=\columnwidth]{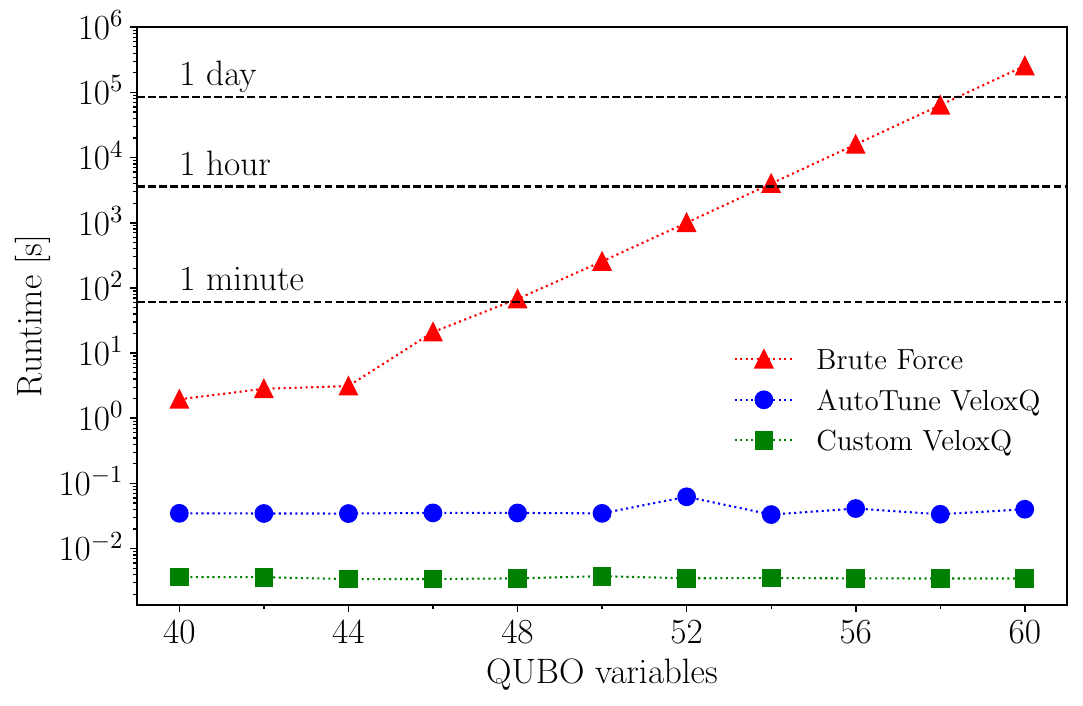}
    \caption{Brute-force runtime plot, reproduced from~\cite{veloxq2025}.}
    \label{fig:original_BF_plot}
\end{figure}

Utilizing expensive 2×4 NVIDIA H100 GPUs, a brute-force solver can handle problems with up to 60 variables in approximately three days~\cite{veloxq2025}, which is equivalent to $\sim 0.56 \times 10^{12}$ states per second on a single NVIDIA H100 GPU. However, as the corresponding work~\cite{jalowiecki2025} is still in preparation, its algorithmic complexity and implementation details are not publicly known. 
Nevertheless, the near-linear trend of $\log\bigl(\text{Runtime [s]}\bigr)$ vs $N$ in Fig.~\ref{fig:original_BF_plot}, implies an $O(1)$ per-state complexity for the underlying algorithm.

Developing faster and more accurate brute-force solvers along with hybrid quantum-classical algorithms represents a promising direction for future research.

\subsection{Our Contribution}
In this work, we present a novel approach to exact QUBO solving based on matrix formalism and prefix-suffix decomposition. 
It naturally maps the exhaustive search algorithm to the massively parallel GPU architecture. 
Gray code ordering and energy components precomputation reduce overall complexity to $O(1)$ per state.
Adjustable energy components decomposition enables aggressive caching of these components directly in GPU register memory. 
This minimizes memory bandwidth requirements and transforms the problem from \emph{memory-bound} to \emph{compute-bound}. 

The proposed CUDA implementation demonstrates an evaluation speed of $7.5\times10^{12}$ states per second, setting a state-of-the-art performance standard for exhaustive solvers.
Furthermore, this implementation achieves tera-scale performance on affordable, low-memory GPUs, making it a cost-effective solution.

\section{Mathematical Framework and Notation}
\label{sec:math_framework}

\subsection {Prefix-Suffix Decomposition Approach}
\label{sec:pref_suff_decompose}

Consider an arbitrary set of $M$ binary state vectors arranged as rows in the matrix $\mathbf{X} \in \{0,1\}^{M \times N}$.
Given an upper-triangular QUBO matrix $\mathbf{Q} \in \mathbb{R}^{N \times N}$,
we compute the vector $\mathbf{E} \in \mathbb{R}^M$ where each element represents the energy of the corresponding state:
\begin{equation}
\mathbf{E} = \text{diag}(\mathbf{X} \mathbf{Q} \mathbf{X}^T).
\label{eq:total_energy}
\end{equation}

To establish the foundation for parallel computation of energy contributions, we introduce the following decomposition.
Each binary state vector $\mathbf{x} \in \{0,1\}^N$ can be represented as a concatenation of two subvectors:
\begin{equation}
\label{eq:state_partition}
\mathbf{x} = [\mathbf{x}_p, \mathbf{x}_s],
\end{equation}
where $\mathbf{x}_p \in \{0,1\}^A$ is the prefix part containing the first $A$ bits, and $\mathbf{x}_s \in \{0,1\}^B$ is the suffix part containing the last $B$ bits, with $A + B = N$.

Following the partitioning~\eqref{eq:state_partition}, we split the state matrix $\mathbf{X} \in \{0,1\}^{M \times N}$ and the QUBO matrix $\mathbf{Q} \in \mathbb{R}^{N \times N}$ into conformal blocks according to prefix size $A$ and suffix size $B$:

\begin{equation*}
\mathbf{X} = [\mathbf{X}_p, \mathbf{X}_s], \quad 
\mathbf{Q} = 
\begin{bmatrix}
\mathbf{Q}_{pp} & \mathbf{Q}_{ps} \\
\mathbf{0} & \mathbf{Q}_{ss}
\end{bmatrix},
\end{equation*}

where $\mathbf{X}_p \in \{0,1\}^{M \times A}$, $\mathbf{X}_s \in \{0,1\}^{M \times B}$, and the submatrices $\mathbf{Q}_{pp} \in \mathbb{R}^{A \times A}$, $\mathbf{Q}_{ps} \in \mathbb{R}^{A \times B}$, and $\mathbf{Q}_{ss} \in \mathbb{R}^{B \times B}$.

\newpage
The block multiplication yields:
\[\mathbf{X} \mathbf{Q} \mathbf{X}^T = 
\begin{bmatrix}
\mathbf{X}_p & \mathbf{X}_s
\end{bmatrix}
\begin{bmatrix}
\mathbf{Q}_{pp} & \mathbf{Q}_{ps} \\
\mathbf{0} & \mathbf{Q}_{ss}
\end{bmatrix}
\begin{bmatrix}
\mathbf{X}_p^T \\ \mathbf{X}_s^T
\end{bmatrix} =
\]

\begin{equation}
= \mathbf{X}_p \mathbf{Q}_{pp} \mathbf{X}_p^T + \mathbf{X}_p \mathbf{Q}_{ps} \mathbf{X}_s^T + \mathbf{X}_s \mathbf{Q}_{ss} \mathbf{X}_s^T.
\label{eq:energy_decomposition_1}
\end{equation}

Taking the diagonal of both sides of equation (\ref{eq:energy_decomposition_1}), we obtain the energy decomposition given by equation (\ref{eq:total_energy}):
\begin{equation}
\mathbf{E} = \mathbf{E}_{p} + \mathbf{E}_{ps} + \mathbf{E}_{s},
\label{eq:energy_decomposition}
\end{equation}

where $\mathbf{E}_{p} = \text{diag}(\mathbf{X}_p \mathbf{Q}_{pp} \mathbf{X}_p^T)$ is the prefix self-energy, $\mathbf{E}_{ps} = \text{diag}(\mathbf{X}_p \mathbf{Q}_{ps} \mathbf{X}_s^T)$ is the prefix-suffix interaction energy, and $\mathbf{E}_{s} = \text{diag}(\mathbf{X}_s \mathbf{Q}_{ss} \mathbf{X}_s^T)$ is the suffix self-energy.

For further use, we define the suffix-dependent energy vector as:
\begin{equation*}
\mathbf{E}_{sd} = \mathbf{E}_{s} + \mathbf{E}_{ps}.
\end{equation*}

The decomposition (\ref{eq:energy_decomposition}) enables independent and parallel computation of energy contributions.

\begin{table*}[htbp]
\caption{Mathematical Notations and Definitions}
\centering
\begin{tabular}{p{0.15\textwidth} p{0.25\textwidth} p{0.5\textwidth}}
\toprule
\textbf{Notation} & \textbf{Entity} & \textbf{Description} \\
\midrule
$N$ & Problem size & Total number of binary variables \\
$A$ & Prefix size & Number of bits in prefix, $A + B = N$ \\
$B$ & Suffix size & Number of bits in suffix, $A + B = N$ \\
$\mathbf{x}$ & State vector & Complete binary state $\mathbf{x} = [\mathbf{x}_p, \mathbf{x}_s] \in \{0,1\}^N$ \\
$\mathbf{x}_p$ & Prefix vector & First $A$ bits of state, $\mathbf{x}_p \in \{0,1\}^A$ \\
$\mathbf{x}_s$ & Suffix vector & Last $B$ bits of state, $\mathbf{x}_s \in \{0,1\}^B$ \\
$\mathbf{X}$ & State sequence matrix & Ordered set of arbitrary $M$ binary states $\{\mathbf{x}_1, \mathbf{x}_2, \ldots, \mathbf{x}_M\}$ represented as matrix $\mathbf{X} \in \{0,1\}^{M \times N}$ \\
$\mathbf{X}_p$ & Prefix sequence matrix & Ordered set of arbitrary  $M$ prefix vectors represented as matrix $\mathbf{X}_p \in \{0,1\}^{M \times A}$ \\
$\mathbf{X}_s$ & Suffix sequence matrix & Ordered set of arbitrary $M$ suffix vectors represented as matrix $\mathbf{X}_s \in \{0,1\}^{M \times B}$ \\
$\mathbf{U}_N$ & Complete state space & Lexicographically ordered set of all possible $2^N$ binary states represented as matrix $\mathbf{U}_N \in \{0,1\}^{2^N \times N}$ \\
$\mathbf{U}_A$ & Prefix state space & Lexicographically ordered set of all $2^A$ prefix configurations represented as matrix $\mathbf{U}_A \in \{0,1\}^{2^A \times A}$ \\
$\mathbf{U}_B$ & Suffix state space & Lexicographically ordered set of all $2^B$ suffix configurations represented as matrix $\mathbf{U}_B \in \{0,1\}^{2^B \times B}$ \\
$\mathbf{Q}$ & QUBO matrix & Upper triangular matrix, $\mathbf{Q} \in \mathbb{R}^{N \times N}$ \\
$\mathbf{Q}_{pp}$ & Prefix block & QUBO submatrix, $\mathbf{Q}_{pp} \in \mathbb{R}^{A \times A}$ \\
$\mathbf{Q}_{ps}$ & Cross block & QUBO submatrix, $\mathbf{Q}_{ps} \in \mathbb{R}^{A \times B}$ \\
$\mathbf{Q}_{ss}$ & Suffix block & QUBO submatrix, $\mathbf{Q}_{ss} \in \mathbb{R}^{B \times B}$ \\
$E(\mathbf{x})$ & State energy & Scalar energy value $\mathbf{x} \mathbf{Q} \mathbf{x}^T$ \\
$\mathbf{E}(\mathbf{X})$ & States energies vector & Energies of states in $\mathbf{X}$, $\mathbf{E} \in \mathbb{R}^M$ \\
$E_p(\mathbf{x}_p)$ & Prefix term & Scalar prefix energy contribution $\mathbf{x}_p \mathbf{Q}_{pp} \mathbf{x}_p^T$ \\
$\mathbf{E}_p(\mathbf{X}_p)$ & Prefix energies vector & Vector of prefix energy contributions for states in $\mathbf{X}_p$, $\mathbf{E}_p \in \mathbb{R}^M$ \\
$E_s(\mathbf{x}_s)$ & Suffix term & Scalar energy contribution $\mathbf{x}_s \mathbf{Q}_{ss} \mathbf{x}_s^T$ \\
$\mathbf{E}_s(\mathbf{X}_s)$ & Suffix energies vector & Vector of suffix energy contributions for states in $\mathbf{X}_s$, $\mathbf{E}_s \in \mathbb{R}^M$ \\
$E_{ps}(\mathbf{x}_p, \mathbf{x}_s)$ & Interaction term & Scalar energy contribution $\mathbf{x}_p \mathbf{Q}_{ps} \mathbf{x}_s^T$ \\
$\mathbf{E}_{ps}(\mathbf{X}_p, \mathbf{X}_s)$ & Interaction energies vector & Vector of interaction energy contributions for state pairs, $\mathbf{E}_{ps} \in \mathbb{R}^M$ \\
$E_{sd}(\mathbf{x}_p, \mathbf{x}_s)$ & Suffix dependent term & Scalar $E_s(\mathbf{x}_s) + E_{ps}(\mathbf{x}_p, \mathbf{x}_s)$ \\
$\mathbf{E}_{sd}(\mathbf{X}_p, \mathbf{X}_s)$ & Suffix dependent energies vector & Vector $\mathbf{E}_s(\mathbf{X}_s) + \mathbf{E}_{ps}(\mathbf{X}_p, \mathbf{X}_s)$ \\
$\mathcal{C}_{\mathbf{x}_p}$ & State space chunk & Lexicographically ordered set of states with fixed prefix $\mathbf{x}_p$, represented as matrix $\mathcal{C}_{\mathbf{x}_p} \in \{0,1\}^{2^B \times N}$ \\
$\mathbf{1}_{2^B}$ & $2^B$-size vector of ones & \\
$\mathbf{M}_{int}$ & Prefix-suffix interaction matrix & Matrix $\mathbf{M}_{int} = (\mathbf{Q}_{ps} \mathbf{U}_B^T)$, size $A \times 2^B$ \\
\bottomrule
\end{tabular}
\end{table*}

\subsection{State Space Partitioning}

Let $\mathbf{U}_N \in \{0,1\}^{2^N \times N}$ denote the lexicographically ordered set of all possible binary vectors of size $N$. This set is represented as a matrix, where each row corresponds to one binary vector, as illustrated in Figure \ref{fig:u_n_matrix}.
\begin{figure}[h]
\centering
\small
\[
\begin{array}{c}
\hspace{-0.7cm}\overbrace{\phantom{\begin{matrix}0 & 0 & \cdots & 0 & 0 & 0\end{matrix}}}^{N} \\
\left(\begin{matrix}
0 & 0 & \cdots & 0 & 0 & 0 \\
0 & 0 & \cdots & 0 & 0 & 1 \\
0 & 0 & \cdots & 0 & 1 & 0 \\
\vdots & \vdots & \ddots & \vdots & \vdots & \vdots \\
1 & 1 & \cdots & 1 & 1 & 0 \\
1 & 1 & \cdots & 1 & 1 & 1
\end{matrix}\right)\kern-\nulldelimiterspace
\left.\vphantom{\begin{matrix}
0 & 0 & \cdots & 0 & 0 & 0 \\
0 & 0 & \cdots & 0 & 0 & 1 \\
0 & 0 & \cdots & 0 & 1 & 0 \\
\vdots & \vdots & \vdots & \ddots & \vdots & \vdots \\
1 & 1 & \cdots & 1 & 1 & 0 \\
1 & 1 & \cdots & 1 & 1 & 1
\end{matrix}}\right\}\scriptstyle 2^N
\end{array}
\]
\caption{The complete state space matrix $\mathbf{U}_N$ with $2^N$ rows and $N$ columns, ordered lexicographically.}
\label{fig:u_n_matrix}
\end{figure}

We define the matrices $\mathbf{U}_A \in \{0,1\}^{2^A \times A}$ and $\mathbf{U}_B \in \{0,1\}^{2^B \times B}$ similarly for sizes $A$ and $B$, respectively. Thus, $\mathbf{U}_A$ contains all possible prefixes and $\mathbf{U}_B$ contains all possible suffixes.

\begin{figure}[h]
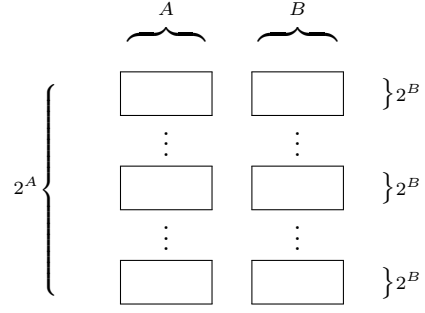

\centering
\small
\[
\begin{array}{cccc}
 & 
\begin{array}{c}
\scriptstyle A \\
\overbrace{\phantom{\,\,\,\,\,\,A\,\,\,\,\,\,\,}} \\
\end{array} & 
\begin{array}{c}
\scriptstyle B \\
\overbrace{\phantom{\,\,\,\,\,\,B\,\,\,\,\,\,\,}} \\
\end{array} &  \\ 
\multirow{5}{*}{$\scriptstyle 2^A \left\{\begin{array}{c}\\ \\ \\ \\ \\ \\ \\ \end{array}\right.$} & \fbox{\phantom{(2  ,  2)}} & \fbox{\phantom{(2  ,  3)}} & \multirow{1}{*}{$\left.\vphantom{\begin{array}{c}\\ \end{array}}\right\} \scriptstyle 2^B$} \\
 & \vdots & \vdots &  \\
 & \fbox{\phantom{(4  ,  2)}} & \fbox{\phantom{(4  ,  3)}} & \multirow{1}{*}{$\left.\vphantom{\begin{array}{c}\\ \end{array}}\right\} \scriptstyle 2^B$} \\
 & \vdots & \vdots &  \\
 & \fbox{\phantom{(6  ,  2)}} & \fbox{\phantom{(6  ,  3)}} & \multirow{1}{*}{$\left.\vphantom{\begin{array}{c}\\ \end{array}}\right\} \scriptstyle 2^B$} \\
\end{array}
\]
\caption{Decomposition of the state space $\mathbf{U}_N$ into $2^A$ chunks of $2^B$ rows, and then into a prefix block and a suffix block.}
\label{fig:u_n_chunks}
\end{figure}

We partition the matrix $\mathbf{U}_N$ into $2^A$ non-overlapping chunks, each containing $2^B$ rows. Further, each chunk is  partitioned into a prefix block of the first $A$ columns and a suffix block of the last $B$ columns, as depicted in Figure~\ref{fig:u_n_chunks}.

\begin{figure}[h]
\centering
\small
\[
\left[
\begin{array}{c|c}
\begin{bmatrix}
\mathbf{x}_p \\
\mathbf{x}_p \\
\vdots \\
\mathbf{x}_p
\end{bmatrix}
= \mathbf{1}_{2^B}^\top \cdot \mathbf{x}_p
& \mathbf{U}_B
\end{array}
\right]
\]
\caption{Chunk structure}
\label{fig:chunk_struc}
\end{figure}

This method of partitioning $\mathbf{U}_N$ exhibits several key properties:

\begin{enumerate}
    \item The suffix block is identical across all chunks and is precisely the matrix $\mathbf{U}_B$.
    
    \item The prefix block within any given chunk consists of identical rows. In other words, each chunk contains all possible suffixes for a single, fixed prefix, as depicted in Figure~\ref{fig:chunk_struc}.
    
    \item For the $i$-th chunk, the common prefix is the $i$-th row of the matrix $\mathbf{U}_A$.
    
    \item A chunk can be uniquely identified by its fixed prefix. We denote the chunk corresponding to prefix $\mathbf{x}_p$ as $\mathcal{C}_{\mathbf{x}_p}$.
    
    \item The prefix block of the chunk $\mathcal{C}_{\mathbf{x}_p}$ can be constructed as the product $\mathbf{1}_{2^B}^\top \cdot \mathbf{x}_p$, where $\mathbf{1}_{2^B} \in \mathbb{R}^{2^B}$ is a vector of ones of size $2^B$.
\end{enumerate}

Thus, the chunk $\mathcal{C}_{\mathbf{x}_p}$ can be expressed by horizontal concatenation as:
\begin{equation}
    \mathcal{C}_{\mathbf{x}_p} = [\mathbf{1}_{2^B}^\top \cdot \mathbf{x}_p, \mathbf{U}_B].
    \label{eq:chunk_structure}
\end{equation}


\subsection{Energy Vector Partitioning}
The complete energy vector $\mathbf{E}(\mathbf{U}_N)$ containing energies of all $2^N$ possible states can be constructed by concatenating the energy vectors of individual chunks:

\begin{equation*}
\mathbf{E}(\mathbf{U}_N) = [\mathbf{E}(\mathcal{C}_{\mathbf{x}_p^{(0)}}), \mathbf{E}(\mathcal{C}_{\mathbf{x}_p^{(1)}}), \ldots, \mathbf{E}(\mathcal{C}_{\mathbf{x}_p^{(2^A-1)}})].
\end{equation*}

By substituting the chunk structure~\eqref{eq:chunk_structure} into the energy decomposition formula~\eqref{eq:energy_decomposition}, we obtain the energy vector for chunk $\mathcal{C}_{\mathbf{x}_p}$:

\begin{equation}\label{eq:chunk_energy_initial}
\begin{split}
\mathbf{E}(\mathcal{C}_{\mathbf{x}_p}) = {}& E_p(\mathbf{x}_p) \mathbf{1}_{2^B} \\
&+ \text{diag}(\mathbf{1}_{2^B}^\top \mathbf{x}_p \mathbf{Q}_{ps} \mathbf{U}_B^T) + \mathbf{E}_s(\mathbf{U}_B),
\end{split}
\end{equation}
where $\mathbf{1}_{2^B}$ is a $2^B$-dimensional vector of ones.

We isolate the prefix-independent part of the second term and denote it as $\mathbf{M}_{\text{int}}$.

\begin{equation*}
\mathbf{M}_{\text{int}} = \mathbf{Q}_{ps} \mathbf{U}_B^T.
\end{equation*}

We can simplify the diagonal operation:

\begin{equation}\label{eq:diagonal_simplification}
\text{diag}(\mathbf{1}_{2^B}^\top \mathbf{x}_p \mathbf{M}_{\text{int}}) = \mathbf{x}_p \mathbf{M}_{\text{int}}.
\end{equation}

Therefore, by substituting~\eqref{eq:diagonal_simplification} into~\eqref{eq:chunk_energy_initial}, the final expression for the chunk energy vector is:

\begin{equation}\label{eq:chunk_energy_final}
\mathbf{E}(\mathcal{C}_{\mathbf{x}_p}) = E_p(\mathbf{x}_p) \mathbf{1}_{2^B} + \mathbf{x}_p \mathbf{M}_{\text{int}} + \mathbf{E}_s(\mathbf{U}_B).
\end{equation}
Note that both $\mathbf{M}_{\text{int}}$ and $\mathbf{E}_s(\mathbf{U}_B)$ are independent of the prefix configuration $\mathbf{x}_p$ and can therefore be precomputed.

\subsection{Permutation Invariance}
\label{sec:reordering_invariance}

The lexicographical ordering of state spaces $\mathbf{U}_A$ and $\mathbf{U}_B$ is not fundamental to the proposed decomposition framework. The framework remains valid under arbitrary reordering of prefix and suffix configurations.

Let $\pi_A$ and $\pi_B$ be arbitrary permutation operators acting on the row indices of $\mathbf{U}_A$ and $\mathbf{U}_B$ respectively, such that $\mathbf{U}_A' = \pi_A(\mathbf{U}_A)$ and $\mathbf{U}_B' = \pi_B(\mathbf{U}_B)$. The permuted state spaces maintain completeness:
\begin{equation*}
    \mathbf{U}_N' = \{[\mathbf{x}_p', \mathbf{x}_s'] : \mathbf{x}_p' \in \mathbf{U}_A', \mathbf{x}_s' \in \mathbf{U}_B'\} = \mathbf{U}_N.
\end{equation*}

Consequently, the complete energy vector $\mathbf{E}(\mathbf{U}_N')$ is a permutation of $\mathbf{E}(\mathbf{U}_N)$, preserving the global minimum:
\begin{equation*}
    \min \mathbf{E}(\mathbf{U}_N') = \min \mathbf{E}(\mathbf{U}_N).
\end{equation*}

The interaction matrix transforms according to the applied permutation of suffix states:
\begin{equation*}
    \mathbf{M}_{\text{int}}' = \mathbf{Q}_{ps} (\mathbf{U}_B')^T = \mathbf{Q}_{ps} \cdot \pi_B(\mathbf{U}_B)^T.
\end{equation*}

Similarly, the suffix energy vector transforms as:
\begin{equation*}
    \mathbf{E}_s(\mathbf{U}_B') = \mathbf{E}_s(\pi_B(\mathbf{U}_B)).
\end{equation*}

Crucially, these transformed quantities retain their prefix-independence and remain suitable for precomputation. The chunk energy computation~\eqref{eq:chunk_energy_final} generalizes to:
\begin{equation*}
    \mathbf{E}'(\mathcal{C}_{\mathbf{x}_p'}) = E_p(\mathbf{x}_p') \mathbf{1}_{2^B} + \mathbf{x}_p' \mathbf{M}_{\text{int}}' + \mathbf{E}_s(\mathbf{U}_B').
\end{equation*}

This property provides flexibility in state space traversal ordering, which can affect computational complexity and parallelization strategies.

\section{Algorithm description}
\label{sec:algorithm_description}

\subsection{Basic brute-force algorithm with O($A \cdot 2^{N}$) Complexity}
\label{sec:basic_bf_description}

The complete brute-force QUBO solver algorithm operates by partitioning the search space into chunks based on prefix configurations. The algorithm consists of two parts: preprocessing and main search loop.

The first one computes reusable energy components with the following steps:
\begin{enumerate}
    \item Compute the all-prefix energy vector \\ $\mathbf{E}(\mathbf{U}_A) = \text{diag}(\mathbf{U}_A \mathbf{Q}_{pp} \mathbf{U}_A^T)$
    \item Compute the all-suffix energy vector \\ $\mathbf{E}(\mathbf{U}_B) = \text{diag}(\mathbf{U}_B \mathbf{Q}_{ss} \mathbf{U}_B^T)$
    \item Construct the prefix-suffix interaction matrix \\ $\mathbf{M}_{int} = \mathbf{Q}_{ps} \mathbf{U}_B^T$
\end{enumerate}

The overall preprocessing complexity is $O(A^2 \cdot 2^A) + O(B^2 \cdot 2^B) + O(A \cdot B \cdot 2^B)$.

The main search loop systematically explores all $2^A$ prefix configurations to identify the global optimum through chunk-based processing. For each prefix configuration $\mathbf{x}_p \in \mathbf{U}_A$, the algorithm:
\begin{enumerate}
    \item Computes the suffix-dependent energy vector $\mathbf{E}_{sd}(\mathbf{x}_p) = \mathbf{x}_p \mathbf{M}_{int} + \mathbf{E}_s(\mathbf{U}_B)$
    \item Finds the local minimum energy and corresponding state index within the chunk using $\mathbf{E}_{sd}(\mathbf{x}_p)$
    \item Adds the prefix energy $E_p(\mathbf{x}_p)$, taken from the precomputed $\mathbf{E}(\mathbf{U}_A)$, to the local minimum to obtain the final energy value
    \item Updates the global minimum if the current local minimum is smaller
\end{enumerate}

The overall search loop complexity is $O(A \cdot 2^B) + O(2^B)$ per prefix configuration, resulting in total complexity of $2^A [O(A \cdot 2^B) + O(2^B)] = O(A \cdot 2^N)$, which dominates over the preprocessing complexity when $A$ (or $B$) are not significantly smaller than $N$. Thus, the total algorithm complexity is $O(A \cdot 2^N)$.

\begin{algorithm}
\caption{QUBO Solver Algorithm}\label{alg:qubo_solver}
\SetAlgoLined
\SetKwInOut{Input}{Input}
\SetKwInOut{Output}{Output}

\Input{QUBO matrix $\mathbf{Q} \in \mathbb{R}^{N \times N}$, partition sizes $A, B$ where $A + B = N$}
\Output{Optimal solution $\mathbf{x}_{opt}$ and minimum energy $E_{min}$}

$\mathbf{E}_A \gets \text{ComputePrefixEnergies}(\mathbf{Q}_{pp})$\;
$\mathbf{E}_B \gets \text{ComputeSuffixEnergies}(\mathbf{Q}_{ss})$\;
$\mathbf{M}_{int} \gets \mathbf{Q}_{ps} \mathbf{U}_B^T$\;
$E_{min} \gets +\infty$\;

\ForEach{$\mathbf{x}_p \in \mathbf{U}_A$}{
    $\mathbf{E}_{cross} \gets \mathbf{x}_p \mathbf{M}_{int}$\;
    $\mathbf{E}_{sd} \gets \mathbf{E}_{cross} + \mathbf{E}_B$\;
    $idx_{local} \gets \arg\min(\mathbf{E}_{sd})$\;
    $E_{local} \gets \min(\mathbf{E}_{sd}) + \mathbf{E}_A[\mathbf{x}_p]$\;
    \If{$E_{local} < E_{min}$}{
        $E_{min} \gets E_{local}$\;
        $\mathbf{x}_{opt} \gets \text{ConstructState}(\mathbf{x}_p, idx_{local})$\;
    }
}

\Return{$(\mathbf{x}_{opt}, E_{min})$}
\end{algorithm}

\subsection{Gray-code optimized algorithm with O($2^{N}$) Complexity}
\label{sec:gray_algorithm}

The key insight is to iterate through all \(2^A\) prefixes in Gray code order~\cite{mutze2024}, where each consecutive element differs by only a single bit. 

\begin{figure}[h]
\centering
\small
\setlength{\tabcolsep}{2pt}
\begin{tabular}{c *{8}{c}}
\toprule
Decimal & 0 & 1 & 2 & 3 & 4 & 5 & 6 & 7 \\
Gray code & 000 & 001 & 011 & 010 & 110 & 111 & 101 & 100 \\
\bottomrule
\end{tabular}
\caption{3-bit Gray code sequence}
\label{fig:gray3}
\end{figure}

Let \(x_p^{(i)}\) and \(x_p^{(i+1)}\) be two such prefixes, differing at position \(k\) (note that \(k\) follows the same left-to-right order as the vector itself).
The prefix change is:
\begin{equation*}\label{eq:prefix_change}
\Delta \mathbf{x}_p = \mathbf{x}_p^{(i+1)} - \mathbf{x}_p^{(i)} = \pm \mathbf{e}_k,
\end{equation*}
where $\mathbf{e}_k$ is the $k$-th standard basis vector and the sign depends on flip direction.

Given that $\mathbf{E}_{sd}(\mathbf{x}_p) = \mathbf{x}_p \mathbf{M}_{int} + \mathbf{E}_s(\mathbf{U}_B)$ from~\eqref{eq:chunk_energy_final}, the change in suffix-dependent energy vector is:
\begin{equation*}\label{eq:energy_change}
\Delta \mathbf{E}_{sd} = \mathbf{E}_{sd}^{(i+1)} - \mathbf{E}_{sd}^{(i)} = \Delta \mathbf{x}_p \mathbf{M}_{int}.
\end{equation*}

Therefore:
\begin{equation*}\label{eq:energy_change_final}
\Delta \mathbf{E}_{sd} = \pm \mathbf{e}_k \mathbf{M}_{int} = \pm \mathbf{M}_{int}^{(k)},
\end{equation*}
where $\mathbf{M}_{int}^{(k)}$ denotes the $k$-th row vector of the matrix $\mathbf{M}_{int}$.

Hence, we can update the energy vector incrementally:
\begin{equation}\label{eq:incremental_update}
\mathbf{E}_{sd}^{(i+1)} = \mathbf{E}_{sd}^{(i)} \pm \mathbf{M}_{int}^{(k)}.
\end{equation}

This reduces the update operation cost from $A \cdot 2^B$ to $2^B$ additions/subtractions.
Such improvement is particularly significant as it simplifies the dominant computational term in the algorithm's overall complexity.

The algorithm starts with zero prefix $\mathbf{x}_p^{(0)} = \mathbf{0}$ and computes initial energy vector $\mathbf{E}_{sd}^{(0)} = \mathbf{E}_s(\mathbf{U}_B)$. For each subsequent prefix in Gray code order:
\begin{enumerate}
    \item Get the flipped bit position $k$ and direction $\delta \in \{-1, +1\}$
    \item Update the energy vector using~\eqref{eq:incremental_update}: $\mathbf{E}_{sd}^{(i+1)} = \mathbf{E}_{sd}^{(i)} + \delta \mathbf{M}_{int}^{(k)}$
    \item Find the local minimum and update the global optimum if needed
\end{enumerate}

For computing prefix energies $\mathbf{E}(\mathbf{U}_A)$ and suffix energies $\mathbf{E}(\mathbf{U}_B)$, the same Gray-code algorithm can be employed. However, instead of searching for minima within each chunk, the energy values are stored to construct the complete energy vectors. This approach reduces the asymptotic complexity of preprocessing from $O(A^2 \cdot 2^A) + O(B^2 \cdot 2^B)$ to $O(2^A) + O(2^B)$. 

Since the main search loop complexity is $O(2^N)$, the preprocessing overhead remains minimal in comparison. Thus, the overall complexity has been reduced from $O(A \cdot 2^N)$ to $O(2^N)$.

\begin{algorithm}
\caption{Gray-code QUBO Solver Algorithm}\label{alg:gray_qubo_solver}
\SetAlgoLined
\SetKwInOut{Input}{Input}
\SetKwInOut{Output}{Output}

\Input{QUBO matrix $\mathbf{Q} \in \mathbb{R}^{N \times N}$, partition sizes $A, B$ where $A + B = N$}
\Output{Optimal solution $\mathbf{x}_{opt}$ and minimum energy $E_{min}$}

$\mathbf{E}_A \gets \text{ComputePrefixEnergies}(\mathbf{Q}_{pp})$\;
$\mathbf{E}_B \gets \text{ComputeSuffixEnergies}(\mathbf{Q}_{ss})$\;
$\mathbf{M}_{int} \gets \mathbf{Q}_{ps} \mathbf{U}_B^T$\;
$E_{min} \gets +\infty$\;
$\mathbf{x}_p \gets \mathbf{0}$\;
$\mathbf{E}_{sd} \gets \mathbf{E}_B$\;

\For{$i = 0$ to $2^A - 1$}{
    $idx_{local} \gets \arg\min(\mathbf{E}_{sd})$\;
    $E_{local} \gets \min(\mathbf{E}_{sd}) + \mathbf{E}_A[\mathbf{x}_p]$\;
    \If{$E_{local} < E_{min}$}{
        $E_{min} \gets E_{local}$\;
        $\mathbf{x}_{opt} \gets \text{ConstructState}(\mathbf{x}_p, idx_{local})$\;
    }
    \If{$i < 2^A - 1$}{
        $(k, \delta) \gets \text{ComputeGrayFlip}(i)$\;
        $\mathbf{x}_p[k] \gets \mathbf{x}_p[k] \oplus 1$\;
        $\mathbf{E}_{sd} \gets \mathbf{E}_{sd} + \delta \cdot \mathbf{M}_{int}[k, :]$\;
    }
}

\Return{$(\mathbf{x}_{opt}, E_{min})$}
\end{algorithm}

\subsection{Parallelization Strategies}
\label{sec:parallelization}
We describe two primary parallelization approaches. 
Both strategies partition the entire search space into segments. 
This enables a map-reduce scheme for global minimum search. 
Each processing unit computes local minima for its assigned segment independently. 
The global minimum is found by reducing the local results.
Furthermore, these two approaches can be combined for hybrid execution.

\subsubsection{Straightforward Parallelization}
\label{subsubsec:simple_parallelization}
Decompose the full search space of size $2^N$ into $2^k$ independent subproblems of size $N - k$.
Split $\mathbf{x} = [\mathbf{x}_f, \mathbf{x}_r]$ with $\mathbf{x}_f \in \{0,1\}^k$, $\mathbf{x}_r \in \{0,1\}^{N-k}$.
Then partition $\mathbf{Q}$ into blocks $\mathbf{Q}_{ff}$, $\mathbf{Q}_{fr}$, $\mathbf{Q}_{rr}$ as in~\ref{sec:pref_suff_decompose}.

For each fixed $\mathbf{x}_f$, form the subproblem by reducing original QUBO matrix as follows:
\[
\mathbf{Q}' = \mathbf{Q}_{rr} + \mathrm{diag}(\mathbf{x}_{f} \mathbf{Q}_{fr}).
\]

For this $\mathbf{x}_f$, the energy becomes
\[
E(\mathbf{x}) = c(\mathbf{x}_f) + E'(\mathbf{x}_r),
\]

where
\[
c(\mathbf{x}_f) = \mathbf{x}_f \mathbf{Q}_{ff} \mathbf{x}_f^\top,
\]\[E'(\mathbf{x}_r) = \mathbf{x}_r \mathbf{Q}' \mathbf{x}_r^\top.
\]

Each of the $2^k$ subproblems is solved independently. The global minimum is obtained over the $2^k$ local minima, each augmented by its corresponding constant $c(\mathbf{x}_f)$.

\subsubsection{Column-wise Parallelization}
\label{subsubsec:column_parallelization}
This approach partitions the suffix space $\mathbf{U}_B$ into multiple segments $\mathbf{U}_B = [\mathbf{U}_{B_1} | \mathbf{U}_{B_2} | \ldots]$. 
Such partitioning induces a corresponding decomposition of the interaction matrix $\mathbf{M}_{int} = [\mathbf{M}_1 | \mathbf{M}_2 | \ldots]$ and energy vector $\mathbf{E}(\mathbf{U}_B) = [\mathbf{E}(\mathbf{U}_{B_1}) | \mathbf{E}(\mathbf{U}_{B_2}) | \ldots]$.
Units share the complete $\mathbf{E}(\mathbf{U}_A)$ vector while storing only their respective $\mathbf{M}_i$ and $\mathbf{E}(\mathbf{U}_{B_i})$ portions.


\section{Implementation Details}
\label{sec:implementation}
We develop two Python-based implementations: a baseline version for demonstration and validation, and a high-performance optimized version designed to achieve maximum performance.


The baseline implementation is available as open-source software through the \texttt{qubo\_lib} library on GitHub~\cite{github_cloud_ru}. 

\subsection{Baseline Implementations}
\label{subsec:baseline}
The core logic is encapsulated within a \texttt{QBF()} solver class, which can be configured for either CPU or GPU execution through NumPy~\cite{numpy2025} and CuPy~\cite{cupy2025} backends respectively.
Both configurations share the same codebase which dynamically selects the appropriate computational backend and memory-management calls at runtime. 
The implementation supports arbitrary data types while delegating overflow handling and parallelism management to the respective libraries.

\subsubsection{NumPy-based CPU Baseline}
\label{subsubsec:cpu}
The NumPy implementation serves as a straightforward, hardware-agnostic reference. By operating on whole vector rows, the algorithm achieves efficiency through cache-friendly memory access patterns and processor vector instructions. While currently operating on a single CPU core, the implementation can be readily adapted to multi-threading through one of the parallelization techniques~\ref{sec:parallelization}.

\subsubsection{Cupy-based GPU Baseline}
\label{subsubsec:gpu-baseline}
The CuPy-based implementation provides a GPU-native execution path while maintaining codebase compatibility with the NumPy variant. It leverages vectorized operations and implicit parallelization across CUDA cores. 
However, the inability to fuse addition and minimum kernels in CuPy leads to multiple memory transfers per iteration, creating a memory-bandwidth bottleneck.
This limitation motivated the development of a custom CUDA kernel.


\subsection{CUDA-based Optimized Implementation}
\label{sec:cuda_impl}

To overcome the limitations of the baseline implementation, we develop a custom CUDA~\cite{cuda2025} kernel-based version of the algorithm. This implementation targets peak brute-force performance.
Our optimization efforts focus on two distinct GPUs: NVIDIA V100 and NVIDIA H100. For clarity, this section focuses on the V100-specific optimizations, while the maximum performance was achieved on the more modern H100.
Many design choices are driven by the characteristics of the target GPU hardware. Below we describe the key optimization approaches:

\subsubsection{Register-Based Energy Components Storage}
\label{subsubsec:register_optimization}
The primary optimization focuses on maximizing the utilization of GPU registers, the fastest available memory tier.
Since we are able to freely choose $A$ and $B$ partition sizes, we can adjust them to make $\mathbf{M}_{int}$ and $\mathbf{E}_{sd}$ small enough. 
Consequently, both components can reside completely in GPU register memory.

Our first target GPU is the NVIDIA Tesla V100 with 80 SMs (streaming multiprocessors) and a 20 MB total register file (256 KB on each SM)~\cite{v100}.
This leads to the optimal suffix size $B=18$.
For $B \geq 19$, insufficient register memory limits the prefix size, and leading to trivial problem sizes. Values of $B \le 17$ underutilized available resources.

\subsubsection{Column-Wise Parallelization}
\label{subsubsec:column_parallelization_impl}

We decompose $\mathbf{M}_{int}$ and $\mathbf{E}_{sd}$ into segments, as shown in section~\ref{subsubsec:column_parallelization}, using CUDA threads as the computational units.
CUDA launch-configuration dimensions are set empirically to 256 blocks and 64 threads per block, creating 16384 total threads.  
This configuration assigns 3.2 threads per FP32 core across the GPU's 5120 (64 per SM) cores, which presumably improves warp-scheduling efficiency.

Column-wise parallelization implies that all threads share the prefix-energy vector. 
This vector is substantial in size (4 GB for $N=49$), requiring storage in GPU global memory. 
However, at a brute-force rate of $1.1 \times 10^{12}$ states per second with $B=18$, only $4.2 \times 10^{6}$ prefix energy fetches per second are required. 
This results in a insignificant memory bandwidth of 8 MB/s, confirming the compute-bound nature of our implementation.

\subsubsection{Coalesced Memory Access Patterns}
\label{subsubsec:memory_coalescing}
Natural lexicographical prefix energy computation conflicts with Gray code traversal. This causes scattered global memory accesses, which cannot be fully mitigated by caching. 
Reordering the energy vector to match Gray code sequence achieves fully coalesced access patterns. This reordering yields substantial performance gains, making the preprocessing overhead worthwhile.

\subsubsection{Data Types}
\label{subsubsec:data_types}
Storing energy components in GPU register memory directly influences data type selection. 
For energy computations, we choose the 16-bit signed integer (int16) data type. 
The integer type for energy eliminates floating-point accumulation errors. The 16-bit sizeof halves memory usage and doubles performance compared to 32-bit types.

For prefix representation we choose signed int32 type, imposing a practical limit of $A \leq 31$.
This allows solving problems up to size 49 ($A=31$, $B=18$), which is still large enough to be computationally challenging.
Unsigned uint32 type is rejected due to slower computation, 64-bit types are unnecessary. 

\subsubsection{Overflow Check Elimination}
\label{subsubsec:integer_arithmetic}

Overflow control is critical for integer types but cannot be performed within the CUDA kernel without performance loss. 
Instead, the kernel assumes external guarantees against overflow. 
Prior to launch, the QUBO problem is solved using a simulated annealing solver. 
The obtained minimum energy provides a lower bound for intermediate energy values during brute-force. 
The upper bound is estimated by solving the negated QUBO matrix. 
If the resulting energy spectrum fits within int16 range, overflow is deemed unlikely. 
This method requires safety margins due to its approximate nature.

\subsubsection{Branch Elimination and Min Reduction}
\label{subsubsec:branch_elimination}
Critical loop optimization focuses on eliminating branching within the innermost computation. 
Arithmetic operations were restructured to avoid conditional addition/subtraction selection. 
Minimum finding was implemented using tree-based reductions with intrinsic \texttt{min()} operations rather than iterative searches.
This provides measurable performance gains through improved instruction throughput.

\medskip
CUDA architecture complexity made analytical parameter selection impractical. 
Final configuration emerged through extensive empirical experimentation, balancing register usage, memory bandwidth, and computational throughput. 
The non-linear optimization process reflected the complex interactions between hardware constraints and algorithm requirements.


\subsection{Scaling Challenges}
\label{subsec:scaling_challenges}

As discussed in~\ref{subsubsec:data_types}, the CUDA kernel is constrained to problem sizes up to $N=49$.
To overcome this limit, we adopt the embarrassingly parallel decomposition introduced in Section~\ref{subsubsec:simple_parallelization}.
Rather than modifying the highly optimized CUDA kernel, the problem structure is adapted to fit the kernel constraint.
We set $|\mathbf{x}_r|$ to $49$, and the remaining $|\mathbf{x}_f| = N - 49$ bits induce a subproblem matrix $Q'$ as defined in~\ref{subsubsec:simple_parallelization}.
For each subproblem, the CUDA kernel finds the local minimum. 
The global minimum is then reduced over all local minima.

Since each node independently constructs $\mathbf{E}(\mathbf{U}_A)$, $\mathbf{M}_{int}$, and $\mathbf{E}(\mathbf{U}_B)$, the approach carries a small computational overhead. We estimate this overhead at under $0.3\%$ of total runtime (Table~\ref{tab:execution_time}).

\section{Experimental Evaluation}
\label{sec:experiments}

\subsection{Setting}
All experiments were conducted on a server running Debian 12 (bookworm) with the Linux kernel 5.15.0-130-generic and Python 3.10.14. 
The experiments were executed within a Docker container running on a virtualized server to ensure a reproducible software environment.

The computational hardware consists of a dual-socket Intel Xeon Gold 6348 CPU~\cite{xeon}, providing 28 physical cores at 2.60 GHz. 
The CPU features a multi-level cache hierarchy with 48 KB L1 data cache and 32 KB L1 instruction cache per core, a 1.25 MB L2 cache per core, and a 39 MB shared L3 cache per socket. 
The processor supports a wide range of SIMD instruction sets, including AVX2, FMA3, and AVX-512~\cite{avx512}, which are leveraged by the numerical libraries. 
The system is equipped with 499 GB of RAM.

The primary computational device is an NVIDIA Tesla V100S-PCIE-32GB GPU, which is based on the Volta architecture~\cite{v100}. 
It features 80 Streaming Multiprocessors (SMs) for a total of 5120 CUDA cores. 
Each SM has access to a 128 KB L1 cache and a shared 6 MB L2 cache. 
The GPU operates at a boost clock of 1597 MHz, delivers a theoretical 16.4 TFLOPS of single-precision (FP32) performance, and provides a memory bandwidth of 1130 GB/s. 

Additionally, an NVIDIA H100 80GB GPU~\cite{h100} based on the Hopper architecture was employed to assess peak performance. This accelerator integrates 132 Streaming Multiprocessors (SMs), amounting to 16896 CUDA cores, with a 256 KB L1 cache per SM and a 50 MB shared L2 cache. Running at a boost clock of 1845 MHz, it offers a theoretical 66.9 TFLOPS of FP32 compute and a memory bandwidth of 3350 GB/s.

The software stack includes NVIDIA driver version 565.57.01, CUDA Toolkit 12.7, and CUDA Runtime version 12.9.
Our implementations leverage NumPy 2.2.6 for the CPU baseline and CuPy 13.6.0 for the GPU baseline and as the interface for our custom CUDA kernel, which was compiled using GCC 12.2.0.
The NumPy-based CPU baseline utilizes the \texttt{scipy-openblas} library (version 0.3.29) for its linear algebra operations, which was compiled with support for SIMD extensions up to AVX-512.



\subsection{GPU Profiling}
We profiled the CuPy baseline and the custom CUDA kernel to identify performance bottlenecks with Nsight Compute~\cite{nsight2025}. The profiling was performed on the V100 GPU only, as we were unable to profile the H100.

The CuPy implementation is fundamentally memory-bound. 
Its inefficiency stems from a poor memory access pattern, specifically repeated global memory accesses. 
For each prefix, the kernel reads the energy vector $\mathbf{E}_{sd}$ and a row of the matrix $\mathbf{M}_{int}$, writes the updated vector back to global memory, and then immediately reads it again for the minimum reduction. 
This costly memory traffic could be mitigated by an effective cache, but the Read-Modify-Write memory access pattern~\cite{kirk2010} induces severe cache pollution, rendering the cache useless.

The profiler confirms this diagnosis, showing a cache hit rate below 1\%. 
This explains the severe memory bandwidth saturation we observe, which makes the implementation profoundly memory-bound. 
For the int16 data type, the main search loop transfers 8 bytes of data per state, totaling 8 TB of memory traffic for a problem with $N=40$. 
This establishes a theoretical lower bound of 8 seconds on the runtime, assuming an optimistic memory bandwidth of 1 TB/s.

The profiler reports an occupancy of 60-80\% for the main kernels.
This is combined with CPU synchronization and the overhead of launching six separate kernels per main loop iteration.
These factors explain the actual execution time of 15.7 seconds for $N=40$.
 
In contrast, the custom CUDA kernel is compute-bound. 
Its primary advantage is the elimination of global memory traffic during the main search loop. 
Profiling confirms this with a negligible memory bandwidth consumption of 87.25 MB/s and a near-perfect cache hit rate.
The kernel achieves 75\% of the theoretical compute throughput. 
The primary reason for not achieving the theoretical peak is the compiler's data placement strategy, which moved data from registers to shared memory, making the number of registers per warp the limiting factor.
As a result, the custom CUDA kernel solves the problem for $N=40$ in just 0.58 seconds, versus the 15.7 seconds taken by the CuPy baseline.

\subsection{Performance Comparison}

\begin{table*}[htbp]
\centering
\caption{State evaluation rate, $10^{12}$ states/sec}
\begin{tabular}{lcccc}
\hline
\textbf{} & \textbf{QUBO‑ES\textsuperscript{a}} & \textbf{QBF\textsuperscript{b}} & \textbf{QBF CUDA\textsuperscript{c}} & \textbf{Bruteforce\textsuperscript{d}} \\ \hline
CPU                         & 0.000053 & 0.0051 & — & — \\ 
Mid‑range GPU: 2080Ti, V100 & 0.066 & 0.072 & 2.33 & — \\ 
High‑end GPU: H100           & — & — & 7.47 & $\sim 0.56$ \\ 
\hline
\\
\textbf{\textsuperscript{a} QUBO-ES bruteforce (tree DFS + Gray code) from~\cite{worktime2020}} \\ 
\textbf{\textsuperscript{b} this work, baseline} \\ 
\textbf{\textsuperscript{c} this work, optimized} \\ 
\textbf{\textsuperscript{d} Bruteforce implementation from~\cite{veloxq2025, jalowiecki2025}, approximated and scaled} \\ 
\end{tabular}
\label{tab:state_rate}
\end{table*}

\begin{table*}[htbp]
\centering
\caption{One-second QUBO size: $\log_2(\text{states/sec})$}
\begin{tabular}{lcccc}
\hline
\textbf{} & \textbf{QUBO‑ES\textsuperscript{a}} & \textbf{QBF\textsuperscript{b}} & \textbf{QBF CUDA\textsuperscript{c}} & \textbf{Bruteforce\textsuperscript{d}} \\ \hline
CPU                         & 25.7 & 32.3 & — & — \\ 
Mid‑range GPU: 2080Ti, V100 & 35.9 & 36.0 & 41.1 & — \\ 
High‑end GPU: H100           & — & — & 42.8 & $\sim 39.0$ \\ \hline
\\
\textbf{\textsuperscript{a} QUBO-ES bruteforce (tree DFS + Gray code) from~\cite{worktime2020}} \\ 
\textbf{\textsuperscript{b} this work, baseline} \\ 
\textbf{\textsuperscript{c} this work, optimized} \\ 
\textbf{\textsuperscript{d} Bruteforce implementation from~\cite{veloxq2025, jalowiecki2025}, approximated and scaled} \\ 
\end{tabular}
\label{tab:log_rate}
\end{table*}

\begin{table*}[htbp]
\caption{Raw execution time (sec)}
\centering
\begin{tabular}{|c|c|c|c|c|c|c|c|}
\hline
\multicolumn{1}{|c|}{N} & \multicolumn{2}{c|}{CPU} & \multicolumn{3}{c|}{Mid-range GPU: 2080Ti, V100} & \multicolumn{2}{c|}{High-end GPU: H100} \\
\cline{2-8}
\multicolumn{1}{|c|}{} & QUBO-ES\textsuperscript{a} & QBF(NumPy)\textsuperscript{b} & QUBO-ES\textsuperscript{a} & QBF(CuPy)\textsuperscript{b} & QBF CUDA\textsuperscript{c} & QBF CUDA\textsuperscript{c} & Bruteforce\textsuperscript{d} \\
\hline
34 & 322.7 & 3.135* (0.053) & 0.269 & 0.571 (0.310) & 0.120 (0.111) & 0.041 (0.037) & — \\
35 & 641.9 & 6.950 (0.055) & 0.518 & 0.775 (0.298) & 0.150 (0.131) & 0.049 (0.043) & — \\
36 & 1289  & 13.28 (0.057) & 1.021 & 1.259 (0.297) & 0.167 (0.130) & 0.051 (0.040) & — \\
37 & 2565  & 26.93 (0.060) & 2.102 & 2.255 (0.315) & 0.205 (0.139) & 0.060 (0.039) & — \\
38 & 5153  & 54.08 (0.063) & 4.082 & 4.176 (0.313) & 0.258 (0.141) & 0.079 (0.042) & — \\
39 & 10265 & 107.5 (0.079) & 8.513 & 7.933 (0.298) & 0.431 (0.195) & 0.110 (0.036) & — \\
40 & 20629 & 211.9 (0.086) & 16.59 & 15.66 (0.298) & 0.575 (0.112) & 0.182 (0.038) & $\sim 15.7$ \\
41 & —     & 434.3 (0.074) & 32.93 & 31.08 (0.311) & 1.049 (0.112) & 0.335 (0.042) & — \\
42 & —     & 872.3 (0.082) & 65.85 & 61.71 (0.299) & 1.970 (0.115) & 0.620 (0.041) & $\sim 22.7$ \\
43 & —     & 1686 (0.080)  & 134.3 & 123.5 (0.299) & 3.841 (0.117) & 1.200 (0.037) & — \\
44 & —     & 3460 (0.081)  & 272.6 & 246.2 (0.299) & 7.59 (0.120)  & 2.372 (0.038) & $\sim 25.0$ \\
45 & —     & 6752 (0.084)  & 525.4 & 491.7 (0.302) & 15.22 (0.127) & 4.756 (0.041) & — \\
46 & —     & 13690 (0.119) & 1053  & 984.2 (0.327) & 30.49 (0.163) & 9.501 (0.036) & $\sim 170$ \\
47 & —     & —             & 2097  & 1967 (0.300)  & 61.49 (0.524) & 19.01 (0.052) & — \\
48 & —     & —             & 4210  & 3930 (0.309)  & 121.3 (0.246) & 37.85 (0.057) & $\sim 540$ \\
49 & —     & —             & 8515  & 7879 (0.335)  & 238.3 (0.487) & 74.36 (0.077) & — \\
50 & —     & —             & 17038 & 15708 (0.296) & 485.7 (2.232) & 151.2 (0.160) & $\sim 2040$ \\
51 & —     & —             & —     & —             & 968.7 (5.256) & 301.3 (0.328) & — \\
52 & —     & —             & —     & —             & 1945 (10.58)  & 604.7 (0.592) & $\sim 8030$ \\
53 & —     & —             & —     & —             & 3877 (21.00)  & 1207 (2.558)  & — \\
54 & —     & —             & —     & —             & 7701 (15.18)  & 2403 (2.623)  & $\sim 32000$ \\
55 & —     & —             & —     & —             & 15464 (29.89) & 4826 (5.095)  & — \\
56 & —     & —             & —     & —             &  —            &  —            & $\sim 128000$ \\
\hline
\multicolumn{8}{l}{\footnotesize{* Time in parentheses denotes precomputation and synchronization overhead.}} \\
\multicolumn{8}{l}{\footnotesize{\textsuperscript{a} QUBO-ES bruteforce (tree DFS + Gray code) from~\cite{worktime2020}}} \\
\multicolumn{8}{l}{\footnotesize{\textsuperscript{b} this work, baseline}} \\
\multicolumn{8}{l}{\footnotesize{\textsuperscript{c} this work, optimized}} \\
\multicolumn{8}{l}{\footnotesize{\textsuperscript{d} Bruteforce implementation from~\cite{veloxq2025, jalowiecki2025}, approximated and scaled}}
\end{tabular}
\label{tab:execution_time}
\end{table*}

We evaluated our QBF solver against leading brute-force algorithms~\cite{worktime2020, veloxq2025, jalowiecki2025} across three hardware tiers: single CPU core, mid-range GPU (RTX 2080Ti/V100), and high-end GPU (H100). We assess both our baseline implementation and the optimized CUDA version.
For the benchmark we use a uniformly random matrix, since the exhaustive search algorithm's execution trace is independent of the specific problem instance.

Since all compared solvers achieve $O(1)$ complexity per state, their runtime scales linearly with the size of the search space, $2^N$. This property makes the \textbf{state evaluation rate} (states/sec) a primary performance metric instead of raw execution times. 

The performance results are summarized in Table~\ref{tab:state_rate}. For a more intuitive understanding of the scale, Table~\ref{tab:log_rate} presents the same data as $\log_2(\text{states/sec})$. This metric directly corresponds to the QUBO problem size solvable in one second. Finally, Table~\ref{tab:execution_time} provides the raw execution times for reference.

Our comparison involves several methodological approximations. 
First, as the work \cite{jalowiecki2025} is unpublished, we rely on approximate values extracted from their performance plot (Fig.~\ref{fig:original_BF_plot}).
The results were obtained on a 2x4 H100 rig; to enable a direct comparison, we linearly normalized these results to a single GPU.
This optimistic scaling assumes perfect parallel performance with zero communication overhead.

Second, we group results from the NVIDIA RTX 2080Ti (4352 CUDA cores, 13.5 TFLOPS) and Tesla V100 GPUs (5120 CUDA cores, 16.4 TFLOPS), justified by their relatively small performance gap (15-20\%).

Third, our CPU benchmark for QBF benefits from the algorithm's inherent vectorizability, enabling the NumPy backend to explicitly use of AVX-512 instructions  for processing 32 integers (16-bit) in parallel. In contrast, the CPU performance of QUBO-ES is reported without details on its use of SIMD instructions, creating an uneven comparison that likely favors our implementation.

Despite approximations introducing a margin of error, these minor inaccuracies do not alter the qualitative conclusions of the comparison.

The high performance of the QBF solver stems directly from its underlying algorithmic design. 
The method is inherently vectorizable for CPUs and maps naturally to the massive parallelism of GPUs, resulting in a compute-bound, cache-friendly implementation.
This synergy between the algorithm and modern hardware architecture pushes the practical boundaries of the exhaustive search method.

\section{Discussion}

\subsection{Why Not Tensor Cores}
A natural question is whether the algorithm can be accelerated using Tensor Cores, which on the V100 GPU offer a theoretical peak of 112.2 TFLOPS compared to 16.4 TFLOPS for standard CUDA cores. 
The answer is likely no.

The high throughput of Tensor Cores is achieved through the HMMA instruction, a 4x4 matrix multiply-add operation ($D = A \times B + C$) counted as 128 FLOPs. 
Our core computational kernel, however, consists of a vector addition followed by a minimum reduction: $\mathbf{E}_{sd} = \mathbf{E}_{sd} + \mathbf{M}_{int}^{(k)}$ and $\min(\mathbf{E}_{sd})$.

While the addition could be mapped to the addition part of HMMA by reshaping vectors into 4x4 matrices, this would only utilize 16 of the 128 FLOPs. 
This reduces the effective throughput to approximately $(16/128) \times 112.2 \approx 14.0$ TFLOPS. 
Furthermore, the subsequent minimum reduction cannot be performed on Tensor Cores and would require a separate pass on standard CUDA cores, introducing additional overhead. 
Therefore, the current CUDA core-based implementation remains the most efficient and suitable architecture for this problem.

\subsection{Future Work}
The proposed approach opens up several promising directions for future enhancement.

First, the current implementation is optimized for dense matrices. 
Adapting the framework for sparse matrices represents another promising avenue. 
This would involve reformulating the matrix operations and data structures to exploit matrix sparsity could yield performance gains for some problems.

The second future direction is supporting floating-point QUBO matrices. 
This requires addressing the challenge of numerical error accumulation inherent in the long sequential incremental Gray-code updates, potentially through mixed-precision or periodic correction schemes.

Third direction involves adapting the solver for newer accelerator architectures, such as Ampere or Hopper. 
However, achieving high performance will require a new round of empirical tuning to re-balance key parameters for the new hardware.
This effort is necessary to unlock the full potential of any future accelerator.

Furthermore, the framework can be extended beyond finding the single ground state to generate the low-energy spectrum. 
This can be achieved by modifying the kernel to maintain a list of the $k$ lowest energies. 
In turn, this provides a more comprehensive benchmark for quantum and heuristic solvers by revealing the structure of the solution landscape.

\section{Conclusion}
This paper presents a novel parallel exhaustive search algorithm for QUBO problems with dense matrices, achieving O(1) complexity per state. 
The approach combines complete state space energy vector prefix–suffix decomposition with Gray code traversal. 
This enables register-based caching, reduces memory bandwidth requirements to negligible levels, and shifts the bottleneck from memory-bound to compute-bound. 
Our custom CUDA implementation reaches a state evaluation rate of $7.5\times10^{12}$ states per second on a single H100 GPU for the int16 data type. 
An even more significant result is $2.3\times10^{12}$ states per second on a low-memory V100 GPU, which is more affordable and accessible. 
This makes the proposed solver a practical tool for researchers.
The results outperform the best known alternatives by more than an order of magnitude, setting a new state-of-the-art for brute-force exact QUBO solvers. 
The baseline source code is publicly available~\cite{github_cloud_ru}; the optimized solver is also provided for experimental access~\cite{cloud_ru}.

\bibliographystyle{unsrt}
\bibliography{references}

\end{document}